\documentclass[]{interact}

\usepackage{epstopdf}
\usepackage{subfigure}

\usepackage{natbib}
\bibpunct[, ]{(}{)}{;}{a}{}{,}
\renewcommand\bibfont{\fontsize{10}{12}\selectfont}

\theoremstyle{plain}

\theoremstyle{definition}

\theoremstyle{remark}

\begin{document}

\title{Evolutions of Individuals Use of Lyon's Bike Sharing System}

\author{
\name{Jordan Cambe\textsuperscript{a,b}\thanks{CONTACT Jordan Cambe. Email: jordan.cambe@ens-lyon.fr} and Patrice Abry\textsuperscript{a,b} and Julien Barnier\textsuperscript{c} and Pierre Borgnat\textsuperscript{a,b} and Marie Vogel\textsuperscript{c} and Pablo Jensen\textsuperscript{a,b}}
\affil{\textsuperscript{a}Univ Lyon, ENS de Lyon, UCB Lyon 1, CNRS, Laboratoire de Physique, F-69342 Lyon, France; \textsuperscript{b}Institut Rh\^{o}nalpin des Systemes Complexes, IXXI, F-69342 Lyon, France; \textsuperscript{c}Univ Lyon, ENS de Lyon, UJM Saint-Etienne, Univ. Lumi\`ere Lyon 2, Centre Max Weber, ENS Lyon, F-69342 Lyon, France}
}

\maketitle
 
\begin{abstract}
Bike sharing systems (BSS) have been growing fast all over the world, along with the number of articles analyzing such systems. However the lack of temporally large trip databases has limited the analysis of BSS users’ behavior in the long term. This article studies the V\'elo'v - a BSS located in Lyon, France - subscribers’ commitment in the long term and the evolution of their usage over time. Using a 5-year dataset covering 121,000 long-term distinct users, we show the heterogeneous individual trajectories masked by the overall system stability. Users follow two main trajectories: about 60\% remain in the system for at most one year, showing a low median activity (47 trips); the remaining 40\% correspond to more active users (median activity of 96 trips in their first year) that remain continuously active for several years (mean time = 2.9 years). This latter class exhibits a relatively stable activity, decreasing slightly over the years. We show that middle-aged, male and urban users are over represented among the 'stable' users. 
\end{abstract}


\section{Introduction}
Bike Sharing Systems (BSS) have been developing rapidly all over the world in the last decades, being now present in more than 500 cities. The number of studies of BSS has followed a similar pattern, focusing on 3 topics : quantifying BSS characteristics, describing users' socio-demographic profiles and evaluating its impacts on environment and public health. 

The automatic recording of BSS activities has allowed a quantitative description of many BSS characteristics: Circadian and monthly activity patterns (see \cite{borgnat_2011,paris}), average speed (\cite{jensen}), patterns of bicycle flows over the cities (see \cite{paris,jensen,borgnat_2011,borgnat_2013}) and influence of weather conditions (\cite{borgnat_2011}). The knowledge derived from these studies, especially on bicycle flows between stations (see \cite{paris,dung}) and the prediction of bike reallocation schedules (\cite{zhang_et_al}), can help the management of station balancing (see \cite{singla,paris,etienne_latifa}), one of the main financial challenges of BSS (\cite{yang_et_al}).

Socio-demographics profiles of BSS users differ generally from the overall cities demographics. Studies carried out in Europe and North America (see \cite{beecham,lda_consulting,Ogilvie,shaheen,fuller_et_al,raux_2017}) have shown that users are more likely to be young, male, with a high level of education and living in the city center.

Finally, several studies have described the impact of BSS policies on environment and public health (see \cite{buehler}). \cite{shaheen2010,shaheen2011} have listed the benefits of BSS: Emission reductions, individual financial savings, physical activity benefits, reduced congestion and facilitation of multimodal transport connections. Yet, other studies question the real impact of BSS on some of the latter. Notably \cite{shaheen} showed the relatively low impact on people favorite mode of transportation. In particular \cite{lit_review,midgley,lda_consulting,buttner} exhibited, for several cities in Europe and Canada the low substitution rates from car usage to BSS. Most BSS riders are indeed people who used to walk and take public transportation.

Among all the research axes cited, questions remain on the commitment of BSS subscribers in the long term. This due to the lack of accurate trip datasets over long periods of time, as mentioned in \cite{lit_review}. Some articles have tried to characterize travel behaviors using surveys, such as \cite{guo_2017,raux_2017}. But the temporal evolution of users has never been deeply investigated. This is the reason why in this article we approach BSS travel behaviors and usage rates under the temporal angle. We address questions related to BSS sustainability, such as : how long do users remain active over the years ? Does their activity increase, decrease or remain stable? Do these trajectories depend on their level of activity? These questions are addressed using a five years long dataset covering about 150,000 long-term distinct users, among which 13,358 have stayed in the system for the whole period. This article follows previous work on Lyon's BSS, \emph{V\'elo'v}, by \cite{vogel} which, using a single year dataset (2011), characterized users according to their intensity and frequency of uses at different time scales (day, week, month and year). This work found 9 classes of users, ranging from 'extreme users', that use V\'elo'v twice a day on average to 'sunday cyclists', who only use the system a few week-ends per year. Using a single year dataset to classify users has however two main limitations. Firstly, there is no way to distinguish between two possible interpretations for a user that appears to be very active from September to December. This could correspond either to (a) someone arriving in town in September that remains very active for the months/years to come or (b) someone who for an unknown reason uses the system only in those months. The second limitation arises from the impossibility to test the stability of users' characteristics over years, which would allow to interpret them as real user properties. For example, do users classified in 2011 as 'sunday cyclists' retain this characteristic over the years? Have they only used V\'elo'v in this way in 2011 or is this pattern a more personal - and stable - use of the system that lasts for longer periods? 

In this paper, the five years long dataset helps drawing a picture of the way people use V\'elo'v over years. After presenting this dataset, we compute in section 3 user classes on our \emph{5 years dataset} using a similar approach to \cite{vogel}. Section 4 studies in detail the evolutions of users' behaviors over the years and shows heterogeneous individual trajectories masked by the overall system stability. 

\section{Dataset} 
The V\'elo'v program started in 2005 in Lyon, France. The V\'elo'v network now has 340 stations, where roughly 4000 bicycles are available. The stations are in the street and can be accessed at anytime (24/7) for rental or return. More information about the history of V\'elo'v and the deployment of stations can be found in \cite{borgnat_2011}. The dataset used in this work records all bicycle trips from 2011/01 to 2015/12 for the V\'elo'v system. It contains more than 38 million trips made by more than 3.8 million users. Each trip is documented with starting and ending times, duration, a user ID code and a tag describing the class of user (year-long subscriber, weekly or daily subscription, maintenance operation, etc). Data are filtered according to the process used in \cite{vogel}, keeping only year-long users and eliminating any anomalies. This leads to a subset of the original population containing 147,354 long-term users. For each person, we count years from the first active day: For example, a user appearing in the records for the first time on March 16$^{th}$, 2011 will end the first adapted year on March $15^{th}$, 2012. To avoid boundary artifacts for users that are active over several years, we stop recording trips at the anniversary date in 2015, even if there are recorded trips later in 2015. 

Note that our elementary unit of analysis is therefore the 'person-year', i.e. the vector of 21 features for each user and each year. One person can therefore appear several times (up to 5) and change group from year to year. One could adopt a different point of view, using persons as the entities and computing a single vector for each of them, averaged over their whole period of activity. This would have two drawbacks: masking the single user trajectories over the years and comparing vectors computed over different periods (from 1 to 5 years). Comparing the third and fourth columns of Table \ref{tab:comp_cldr_adapt} shows that using the 'person-year' or the 'person' as the basic entity leads to roughly the same proportions for the different classes. We then retain the 'person-year' description, which allows studying users' trajectories.

\section{Classes of users}

In this section, we compute users classes on our 5 years dataset, using a similar approach to \cite{vogel}, and offer a brief description of them.

\subsection{Computing users classes}
From this dataset, we compute the same 21 normalized features characterizing the activity as in \cite{vogel}. For each person, these features quantify the intensity and regularity of use over the year (14 features) and the week (7 features).

\begin{itemize}

\item $trips\ week$, averaged number of trips made per week, calculated over the weeks during users traveled at least once, and normalised dividing by 1.5 times the interquartile range of the distribution for all users (equal to the difference between the lower and upper quartile of the distribution).

\item $trips \ day 1-5$, number of trips per week day. Days are ranked from one to five, \emph{day 1} being the day with the highest number of trips and \emph{day 5} the day with the lowest number of trips. $trips \ saturday$, average number of trips made on Saturdays. $trips \ sunday$, average number of trips made on Sundays. These seven features are normalized over a total sum unity over the week.

\item $trips \ year$, total number of trips made over the adapted year, normalized dividing by 1.5 times the interquartile range of the distribution for all users (i.e. the difference between the lower and upper quartile of the distribution).

\item $trips \ month 1-12$, number of trips per month, normalized to a total sum unity, months are ranked from one to twelve, \emph{month 1} being the month with the highest number of trips and \emph{month 12} the month with the lowest number of trips. 
\end{itemize}

\begin{figure}
\centering
	\subfigure{
    \centering
    \resizebox*{10cm}{!}{\includegraphics{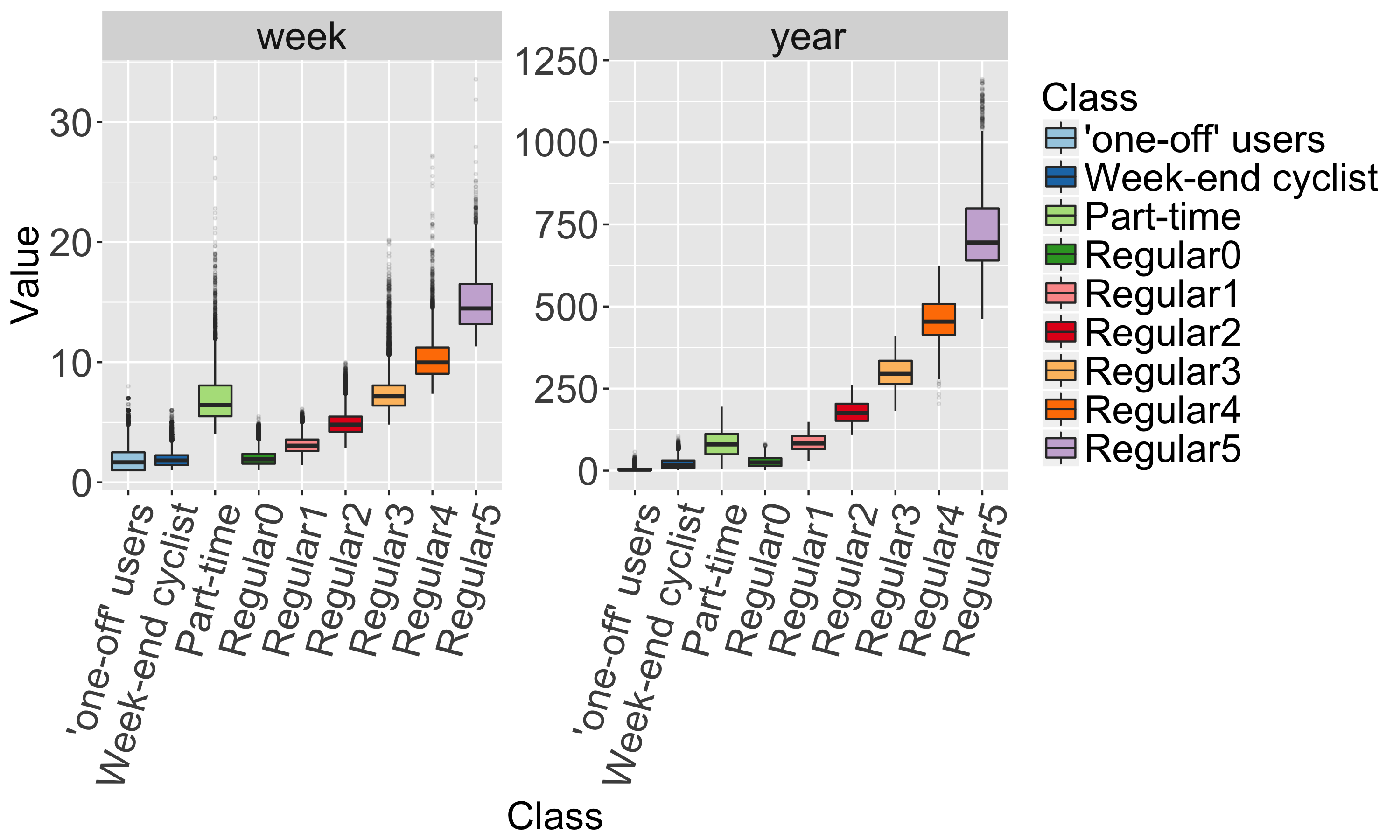}}
    \label{fig:behavk9a}
	}

	\subfigure{
	\resizebox*{10cm}{!}{\includegraphics{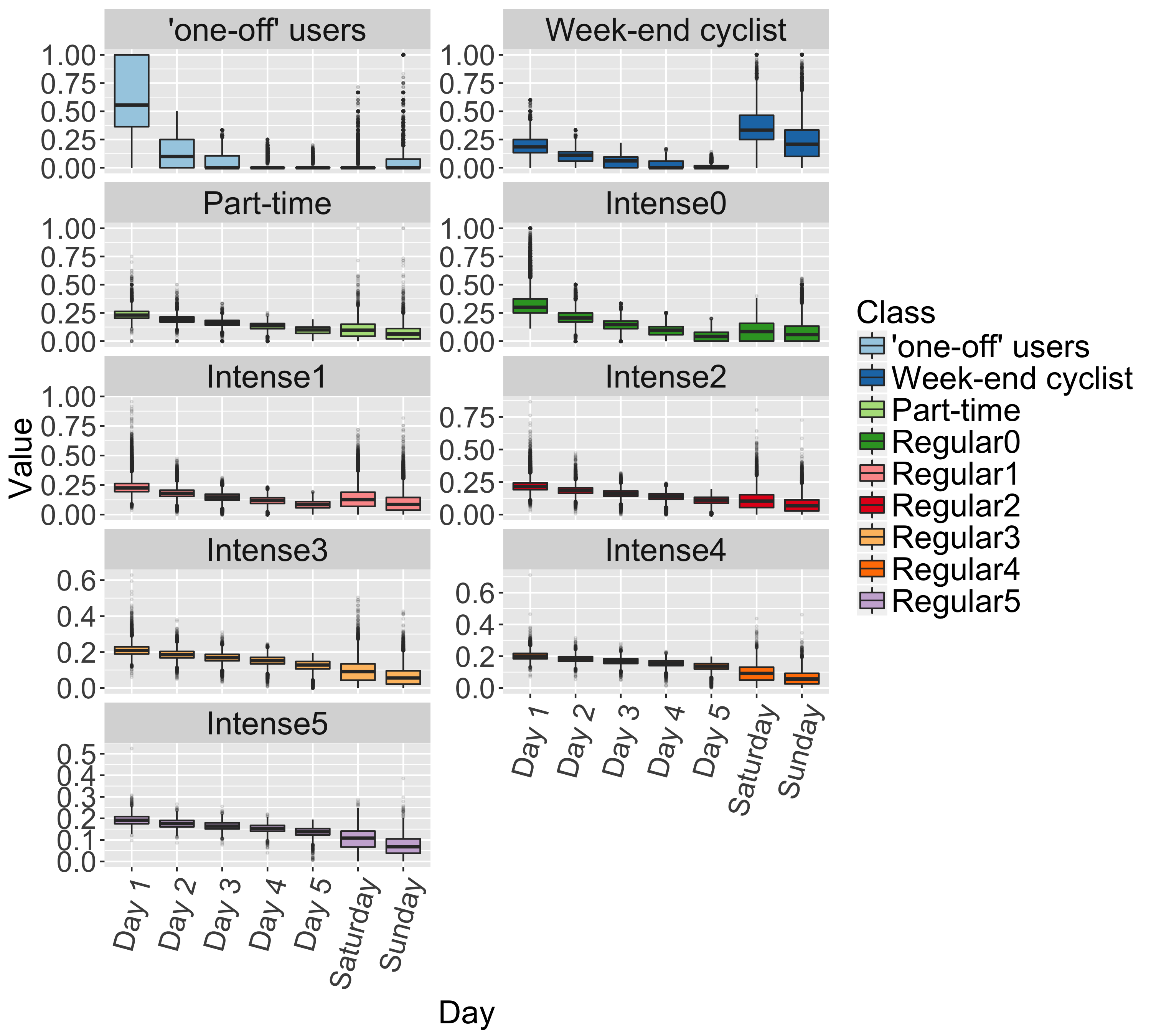}}
	\label{fig:behavk9b}
    }
    
    \subfigure{
	\resizebox*{10cm}{!}{\includegraphics{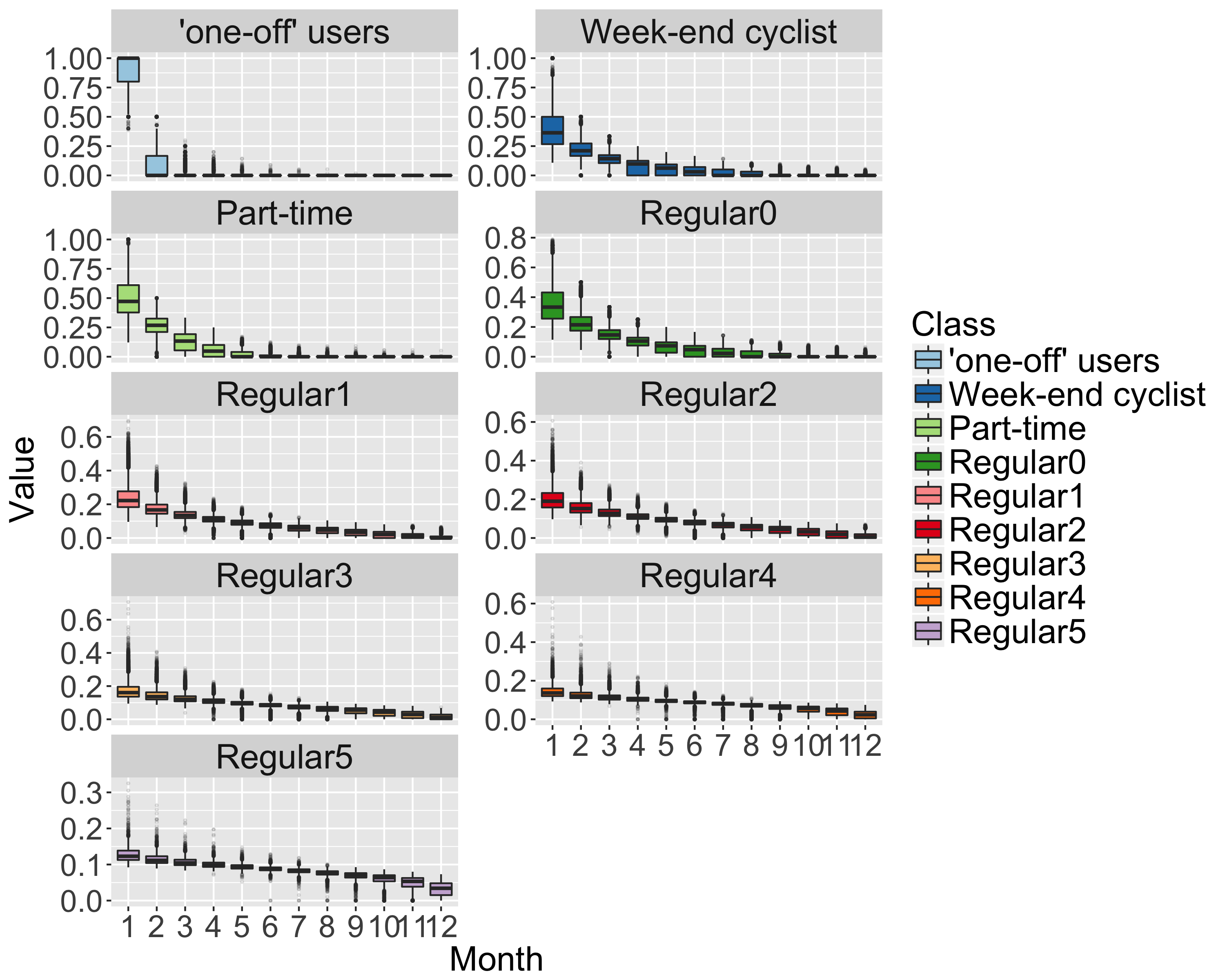}}
	\label{fig:behavk9c}
    }
    
\caption{Boxplots of the behavioral patterns at different time scales of the 9 classes. (a) number of uses per year (right) and per \emph{active} week (left), for each class. A week is considered 'active' for a user whenever he/she takes a bicycle at least once. (b) normalized number of uses for each day of the week and for each class. Week days range from one to five in decreasing order of activity for each user. Saturday and Sunday are computed separately as users' activity is different on week-ends. (c) normalized number of uses for each month of the year and for each class.}
\label{fig:behavk9}
\end{figure}

To allow comparisons with our previous study \cite{vogel}, we use a k-means procedure with the same number of groups (9) to find a partition of users. Robustness tests have shown that there is no other clear partition of our data. A detailed description of the 9 classes is given in Table  \ref{tab:comp_cldr_adapt} and Figure \ref{fig:behavk9}.

\begin{table}
\tbl{Description of user classes found by the k-means for the 21 features. Note that since the entity is a 'person-year', these counts do not directly represent proportions of individuals, because users that stay in the system for long periods are over-represented. However, the comparison with the proportions obtained for year one (third column), which correspond to real users, shows that this effect is relatively weak. \emph{\#trips/year} is the median number of trips in a year for each class.}
{\begin{tabular}{rrrrr}
\hline
Class & \# person-year & freq & $1^{st}$-year freq &  \#trips/year \\
\hline
'one-off' users & 12164 & 5.8 & 5.2 & 3.0\\
Week-end cyclist & 24313 & 11.6 & 11.8 & 17\\
Part-time & 7639 & 3.6 & 4.3 & 80\\
Regular0 & 56849 & 27.1 & 27.1 & 25\\
Regular1 & 51446 & 24.5 & 24.9 & 83\\
Regular2 & 29225 & 13.9 & 14.2 & 175\\
Regular3 & 17183 & 8.2 & 8.0 & 295\\
Regular4 & 8444 & 4.0 & 3.6 & 454\\
Regular5 & 2361 & 1.1 & 0.9 & 695\\
\hline
\label{tab:comp_cldr_adapt}
\end{tabular}}
\end{table}

\subsection{User Classes}
The 9 classes correspond to different profiles of use. There are 6\% of 'one-off' users, who make on average only 3 trips per year, generally the same month and then disappear from the database. Another almost 12\% of users are mainly active in week-ends, either for shopping (Saturdays) or leisure (Sundays) (second line of Table \ref{tab:comp_cldr_adapt}). The last 6 lines of Table \ref{tab:comp_cldr_adapt} represent users that show a regular activity over the year and differ mainly by their intensity of use, from twice a month (regular0 class, gathering 27\% of users) to nearly twice a day (regular5, 1\% of users). The part-time class is quite peculiar: We will show below that it can be interpreted as the class where users end up for the last year of activity.

\section{Evolution of users' behaviors}
The nine classes capture users' activity patterns over the week and the year. The original \emph{5 years database} gives new insights into the evolutions of BSS usage for specific users for more than one year. This section answers questions such as: How long are users likely to remain active? Does this depend on their age, sex, residence or class? What are the differences between stable users, i.e. those that remain in the system for several years and maintain a constant activity and those that use it for several days or months and then quit?

\subsection{Overall evolution}
We first analyze the global system evolution by computing the number of active users, the number of trips and the average number of trips per user for each civil year. Table \ref{tab:general_trend} shows that there is a steady increase in the number of users and trips. However, the average number of trips per user oscillates around 93 trips/year, with no clear long-term trend.

\begin{table}
\tbl{Number of trips per active user.}
{\begin{tabular}{rrrrrr}
\hline
Year & 2011 & 2012 & 2013 & 2014 & 2015 \\
\hline
Active Users & 50,366 & 55,864 & 61,766 & 70,052 & 76,499 \\
Trips & 4,738,197 & 5,187,575 & 5,629,564 & 6,679,906 & 7,098,415 \\
\textbf{Trips per user} & \textbf{94.0} & \textbf{92.8} & \textbf{91.1} & \textbf{95.3} & \textbf{92.7} \\
\hline
\end{tabular}}
\label{tab:general_trend}
\end{table}

\subsection{Individual evolutions}
\begin{figure}
\centering
\subfigure{
	\resizebox*{15cm}{!}{\includegraphics{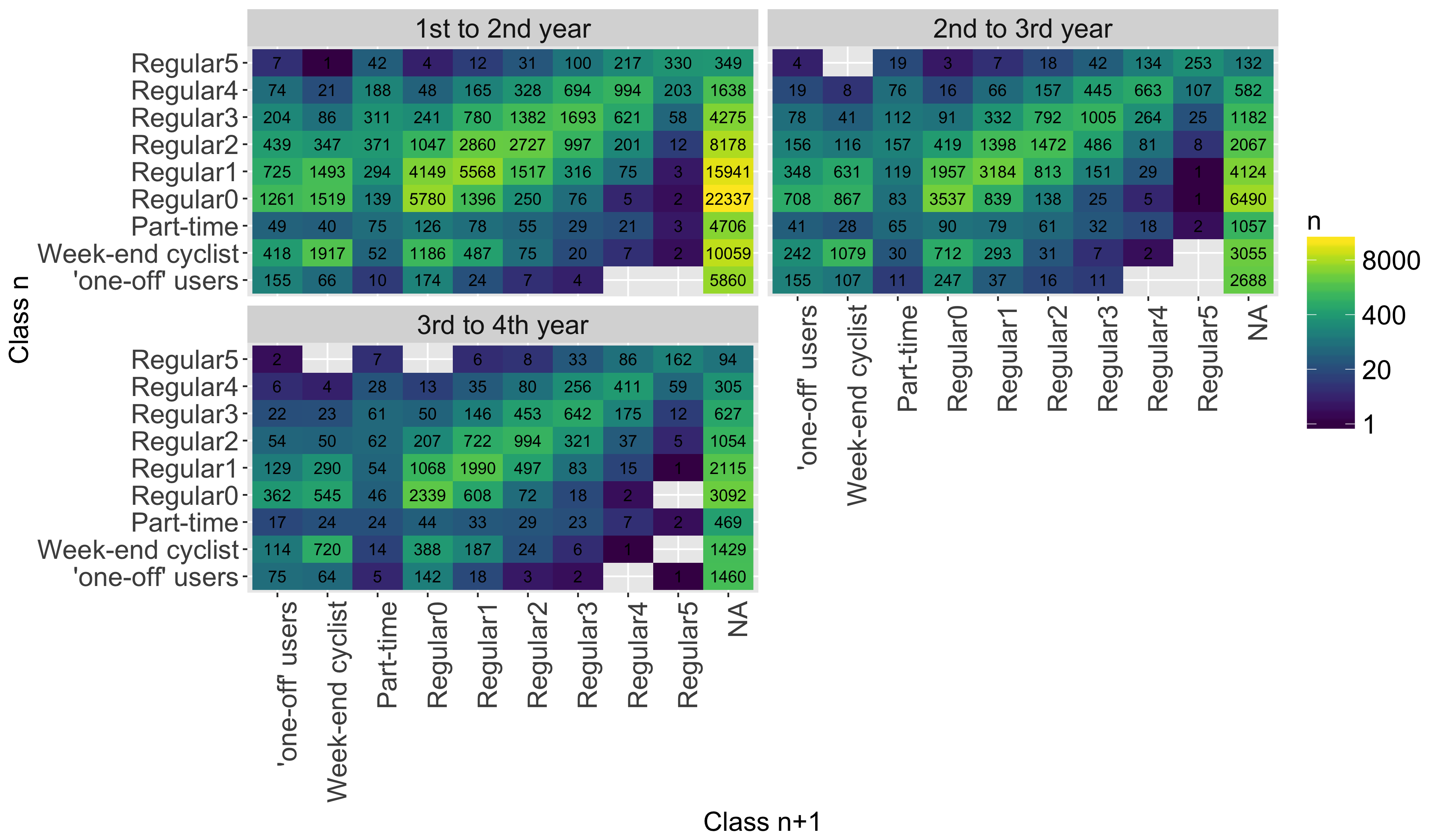}}
    \label{fig:comp_mat_tempb}
	}

\subfigure{
	\resizebox*{15cm}{!}{\includegraphics{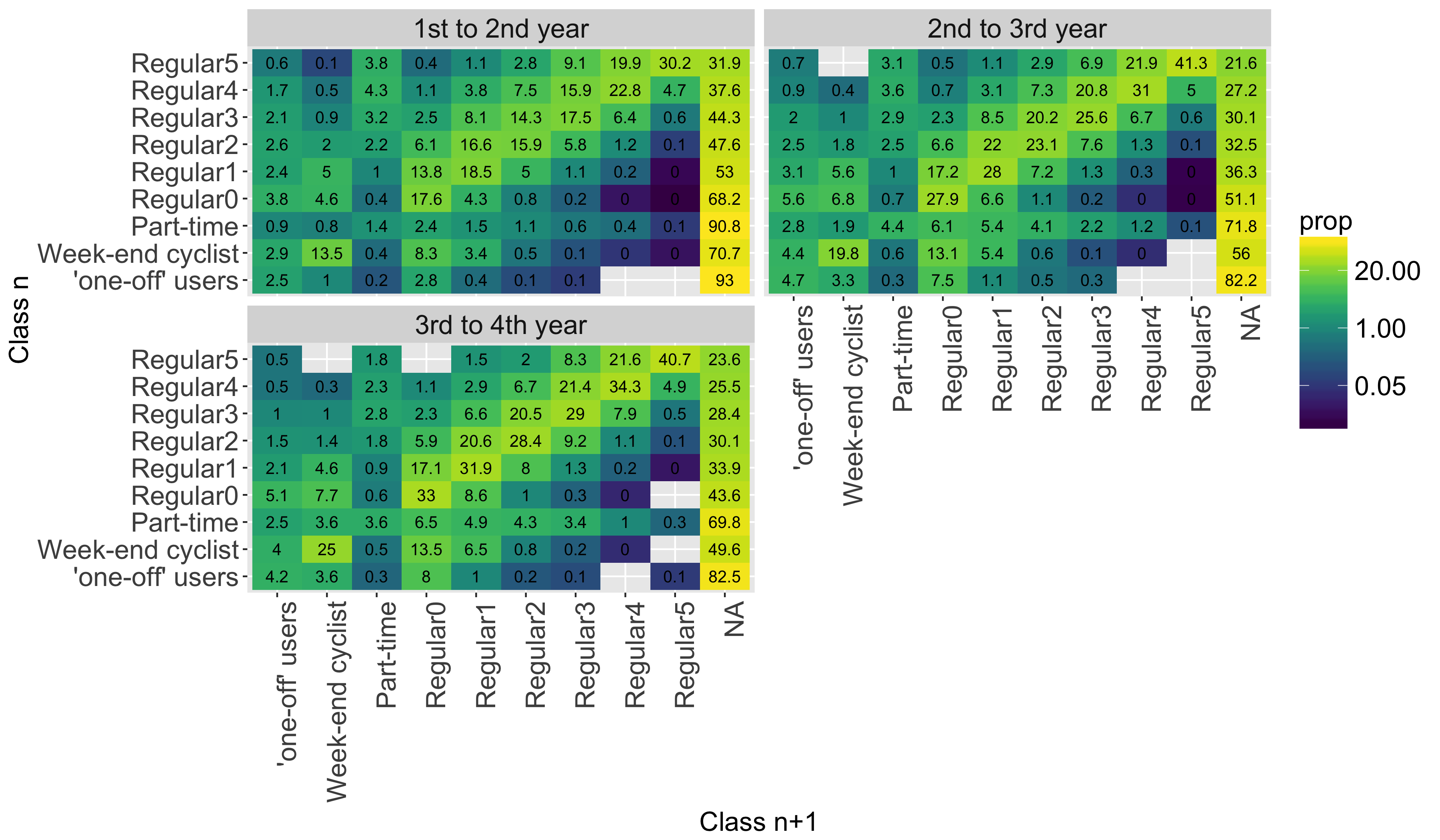}}
    \label{fig:comp_mat_tempa}
	}
\caption{Transfer matrices from year n (lines) to year n+1 (columns). (a) Absolute number of users. Reading: (first line, first table) : 217 users that belonged to the regular-5 class in their first year became regular-4 users in their second year; 330 remained regular-5 and 349 were no longer active; (second line, third table): 411 users that belonged to the regular-4 class in their third year remained regular-4 users in their fourth year; 59 became regular-5 and 305 left the system. (b) Percentage of users. Reading : (first line, first table) : 19.9\% users that belonged to the regular-5 class in their first year became regular-4 users in their second year; 30.2\% remained regular-5 and 31.9\% were no longer active; (second line, third table): 34.3\% users that belonged to the regular-4 class in their third year remained regular-4 users in their fourth year; 4.9\% became regular-5 and 25.5\% left the system.}
\label{fig:comp_mat_temp}
\end{figure}

Although the system is rather stationary at the global level, it is worth noting that individual users exhibit a broad set of temporal behaviors: Some leave simply V\'elo'v and are replaced by new users, some alter their behavior and change class, and many stay in the same class. We start by quantifying the flow of users between classes in consecutive years. Figures \ref{fig:comp_mat_tempa} and \ref{fig:comp_mat_tempb} present these 'transfer matrices' for different years for each class.

These matrices give key informations about the evolution of the system: its overall stability (diagonal terms); the decrease of activity for users that remain active (asymmetry of the non diagonal terms); the high proportions of users leaving it (last column). The following paragraphs give details about these features.

\subsubsection{Most users leave the system after one year}
\begin{figure}
\centering
\resizebox*{12cm}{!}{\includegraphics{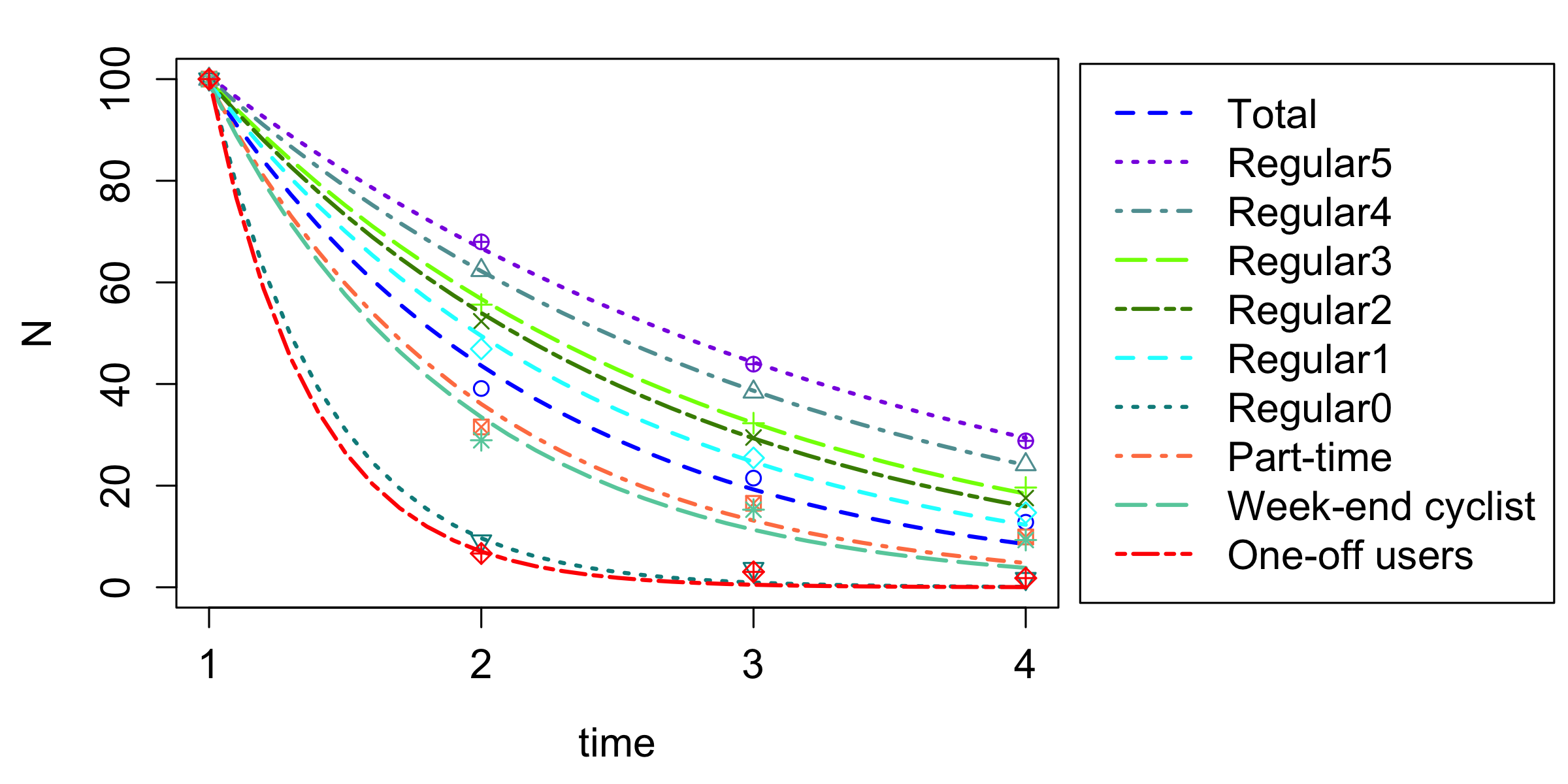}}
\caption{Proportion of users that remain active over adapted years 1 to 4, as a function of their initial class (class observed for their first year). Dashed lines are exponential fits of equation $N_{active}(t) = N_{active}(t=0) \exp(- \frac{1}{\tau} t)$ for each class. For \emph{all users} (long dashed blue line) $\tau = 1.23$. The $\tau$ for each class are given in Table \ref{tab:prop_slopes}.}
\label{fig:leave}
\end{figure}

\begin{figure}
\centering
\resizebox*{12cm}{!}{\includegraphics{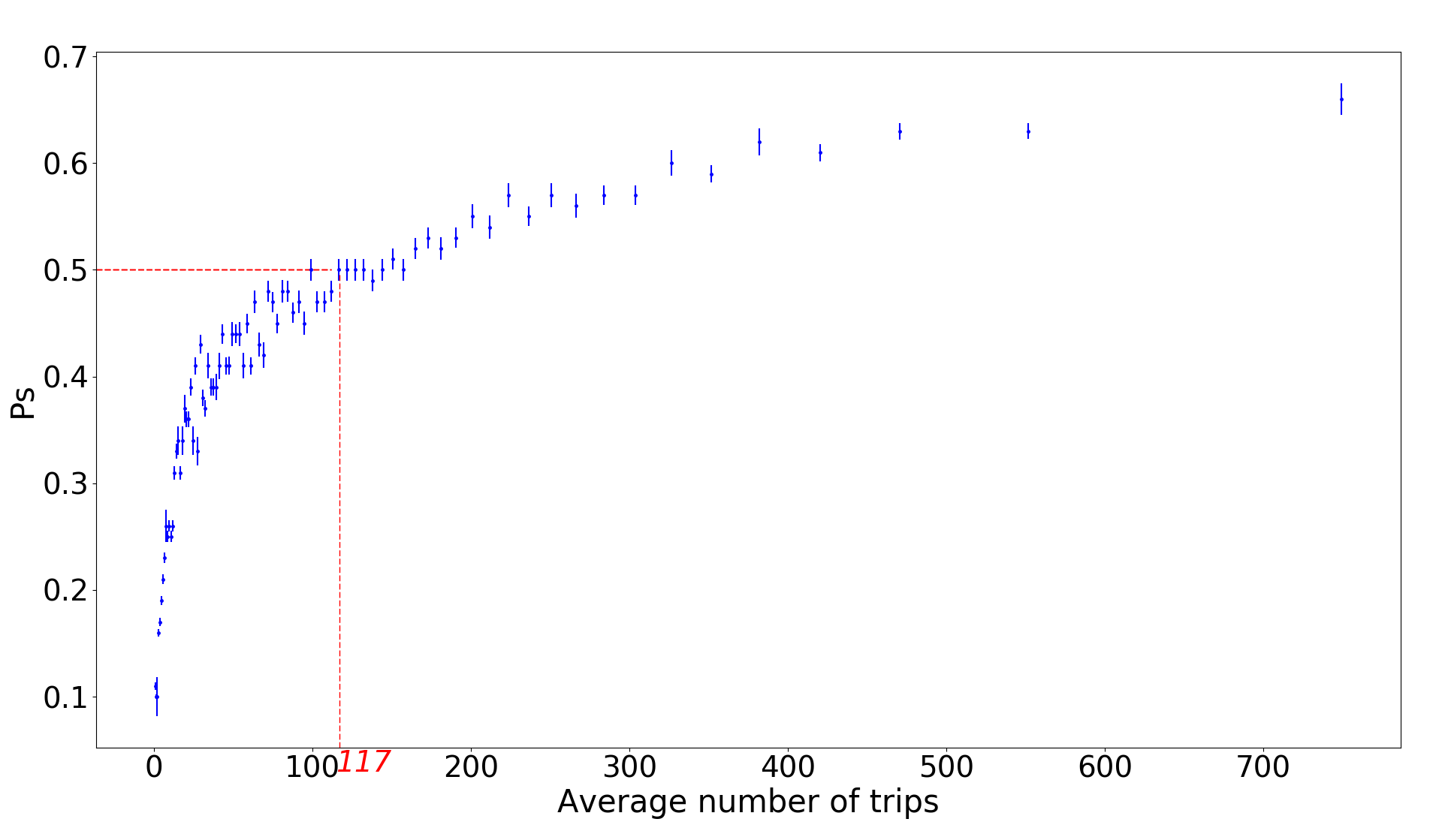}}
\caption{Probability to stay in the system at the end of an adapted year ($P_s$) as a function of the average number of trips during that year. When activity reaches 110-120 trips per year, users are more likely to stay ($P_s \geq 0.5$). Each point represents an average of $P_s$ over 2500 person-years.}
\label{fig:psyears}
\end{figure}

Figure \ref{fig:leave} shows that around 60\% of users stop using Velo'v at the end of each year, the percentage strongly depending on the user class: only 31.9\% of the most intense users, but 93\% of 'one-off' users (Figure \ref{fig:comp_mat_tempa}). Figure \ref{fig:psyears} confirms that, the higher the intensity of use, the higher the probability $P_s$ to stay in the system. 

It is worth noting however that this figure of 60\% might be slightly overestimated. The reason is that users are identified through the ID of one long-term card, the most common being Velo'v own card (35.0\% of the users), public transportation card (Tecely, 55.7\%) and train card (Oura, 5.2\%). The point is that the Tecely cards have to be renewed every 5 years. In some (uncontrolled) cases, this leads to a change of ID, which our analysis interprets as if the user had left the system and another had entered it. To estimate an upper bound on the proportion of incorrectly labeled exits from the system, we note that only 52\% of Velo'v cards users give up after one year, the corresponding figure being 65\% for Tecely users. As the Velov cards do not go through this renewal process, we may estimate that our figure of 60\% users leaving the system at the end of their first year might be overestimated by at most 5-10\%. This corresponds to the proportion of Tecely cards that we expect to be renewed after one year, sensibly lower than 20\% (i.e. 1/5) as many users buy their Tecely card when entering the Velo'v system.

\subsubsection{Users that remain in the system do not change much their behavior \ldots}
Let us now focus on users that remain active for more than one year. The high values found around the matrix diagonals (light colors in Figs \ref{fig:comp_mat_tempa} and \ref{fig:comp_mat_tempb}) show that many users remain in the same class over several years : the probability to stay in the same class is higher than any other probability (except leaving). As discussed briefly below, the second highest probability corresponds to a shift to the neighboring class with lower activity. The matrix also shows that this 'class fidelity' is correlated to the intensity of use: For example, on the first year $P_{I5 \rightarrow I5} = 30.2\% > P_{I4 \rightarrow I4} = 22.8\% > P_{I3 \rightarrow I3} =17.5\%$ \ldots This intensity is therefore a good predictor of future behavior : Staying in the same class or, as discovered earlier (Figure \ref{fig:psyears}), leaving the system. 

\subsubsection{\ldots but they generally reduce their activity}
\begin{figure}
\centering
\resizebox*{12cm}{!}{\includegraphics{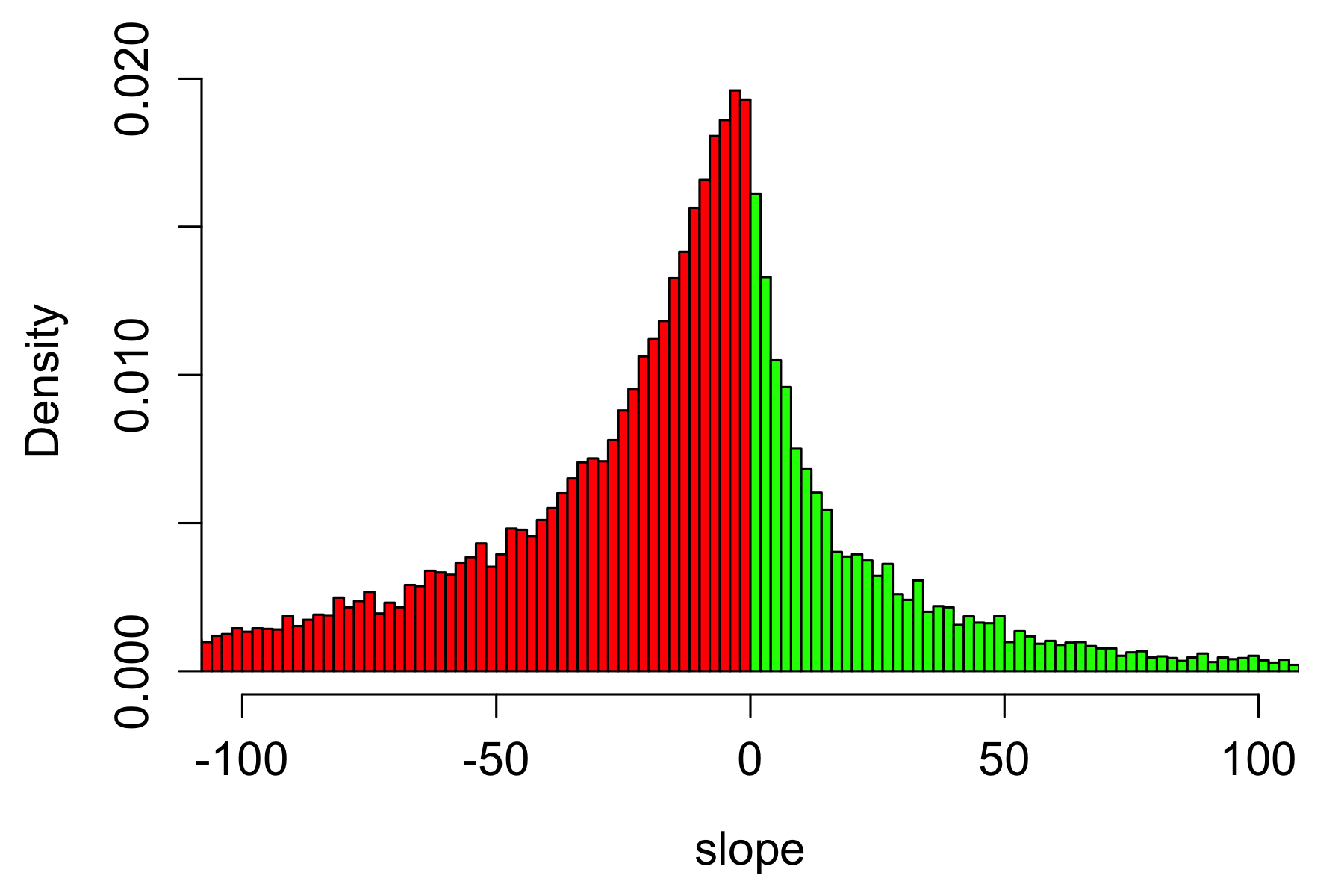}}
\caption{Density distribution of linear regression slopes fitting the number of trips/year over time. Slopes are computed for each user that remained active for at least 3 years.}
\label{fig:hist_slope}
\end{figure}

A careful examination of the matrices shows that they are asymmetrical, the upper part containing higher values than the lower part. As the lines in the graph are ordered by intensity, this means that users have a high probability of reducing their activity from one year to another. To confirm this observation, we studied the individual evolutions of use intensity. For the 25,963 users active for at least 3 years, we obtained their number of trips per year over time. For each user, we computed the slope of the linear regression of their number of trips versus the years, a positive slope meaning an overall increase of use and a negative one a decrease. Figure \ref{fig:hist_slope} shows a plot of the distribution of slopes for all users. As can be seen, it is clearly negatively skewed. Table \ref{tab:prop_slopes} shows that more than two thirds of users reduce their activity over the years, the precise percentage again depending on the classes. 

\begin{table}
\tbl{Influence of the initial (first year) class on user evolution. \emph{Total} : total number of slopes per class. \emph{\% negative} : percentage of negative slopes. $\tau$ : characteristic active time before quitting the system (in years, see Figure \protect\ref{fig:leave}).}
{\begin{tabular}{rrrrr}
\hline
Class & Number & \% negative & $\tau$\\
\hline
Regular5 & 480 & 86.3 & 2.5\\
Regular4 & 1674  & 82.2 & 2.12\\
Regular3 & 3116  & 78.6 & 1.78\\
Regular2 & 5063 & 78.0 & 1.63\\
Regular1 & 7658 & 71.5 & 1.44\\
Regular0 & 5416 & 60.8 & 0.99\\
Part-time & 190 & 61.6 & 0.42\\
Week-end cyclist & 2172 & 56.8 & 0.92\\
One-off users & 191 & 26.2 & 0.37\\
\textbf{Total} & \textbf{25963} & \textbf{70.7} & \textbf{1.23}\\
\hline
\end{tabular}}
\label{tab:prop_slopes}
\end{table}

\subsubsection{And they are older, more likely men and more urban than those who quit}
Comparing the users that remain active for 3 years or more ('stable users') to those that leave after a single year reveals interesting facts about their specific characteristics. Stable users are older (median age 36, against 27, p-value $< 10^{-16}$) and more likely men (men proportion 61\% against 52\%, p-value $< 10^{-16}$), the differences being similar for all the classes. Users that live within the Lyon Villeurbanne perimeter (i.e. the most central districts) are also more likely to stay (40\% against 36\%, p-value $< 10^{-11}$). The differences are even higher for more stable users, i.e. users that remain in the system \emph{and} in the same class for 3 or more years. For example, the median age of the 2,312 regular0 users that remain in the regular0 class is 41.0, instead of 38.0 for the 1,997 regular0 users that remain active but leave the regular0 class (p-value $< 10^{-16}$). The median age of the 10,016 regular0 users that leave the system after their first year is much lower: 28 years.

\subsubsection{Length of use and activity}

\begin{figure}
\centering
\resizebox*{12cm}{!}{\includegraphics{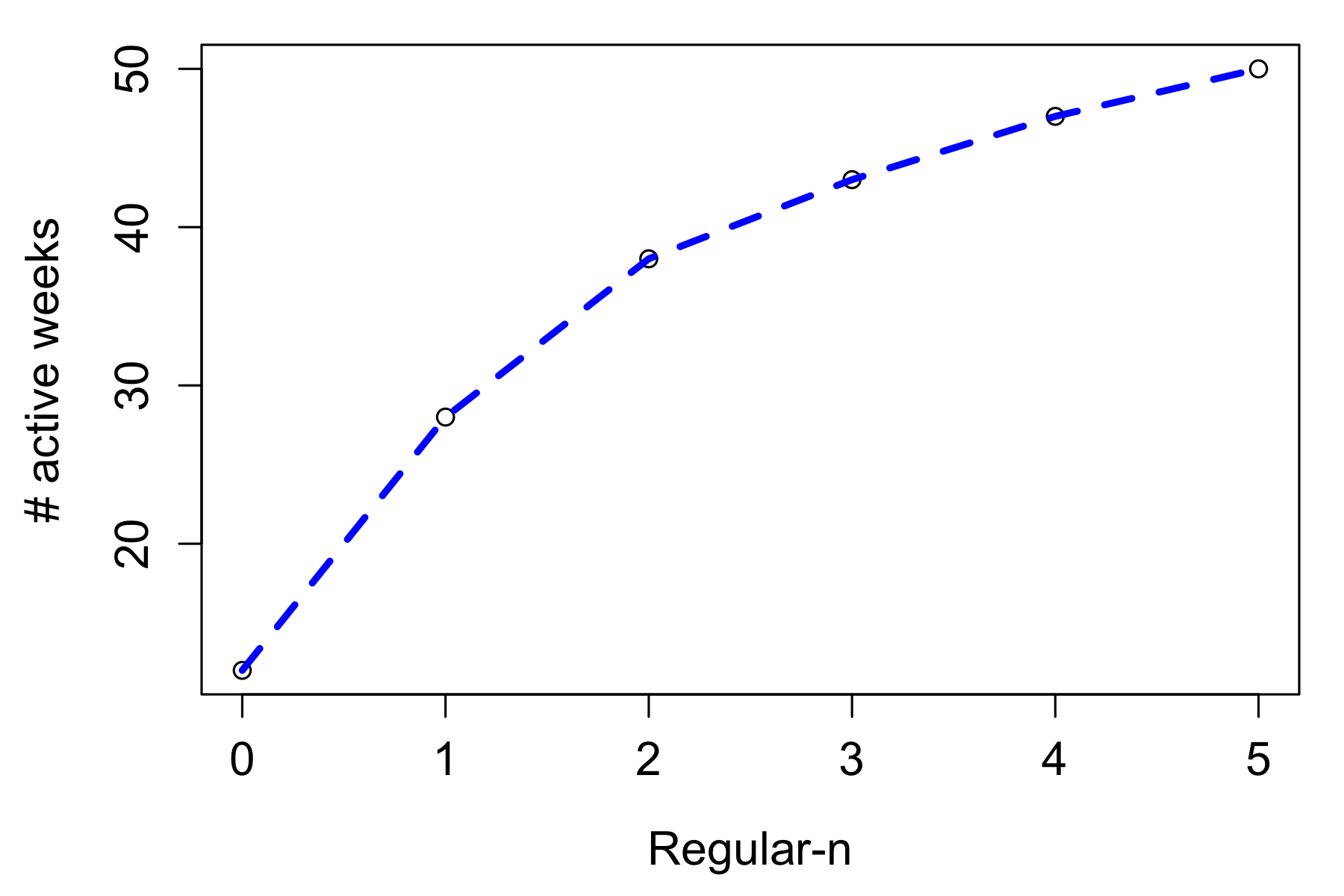}}
\caption{Number of active weeks per year for each class of regular users. An active week is defined as a week where at least a trip took place.}
\label{fig:active_weeks}
\end{figure}

\begin{figure}
\centering
\resizebox*{12cm}{!}{\includegraphics{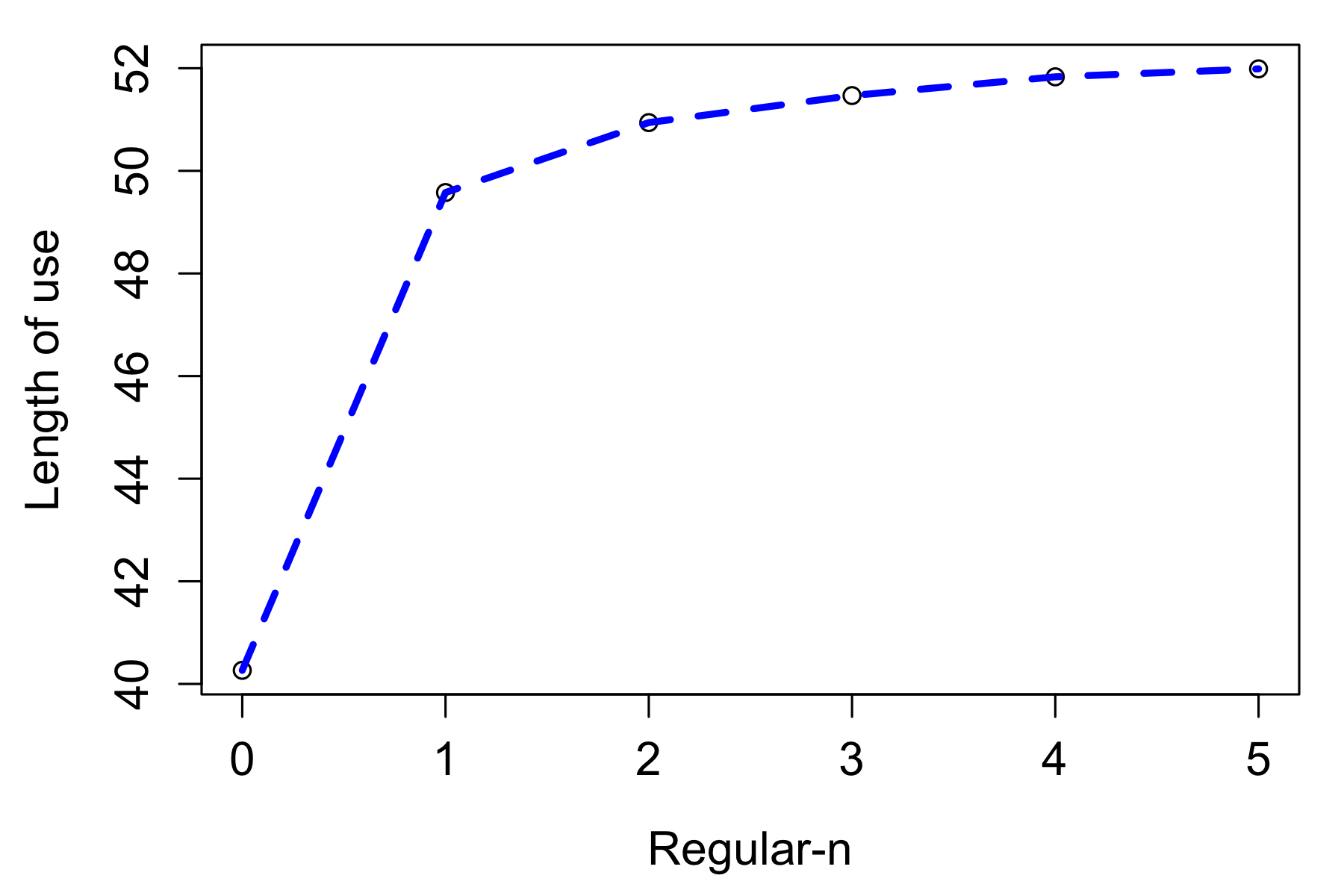}}
\caption{Length of use per adapted year for each class of regular users. The length of use is the time difference between the first active week and the last active week of the adapted year.}
\label{fig:length_use}
\end{figure}

The increase of use from classes regular0 to regular5 arises from the combination of two factors: Both the number of active weeks (figure \ref{fig:active_weeks}) and the number of trips per active week increase. For example, regular5 users are almost 30 times more active over the year than regular0 users (695 instead of 25 trips, see Table \ref{tab:comp_cldr_adapt}), because they are at the same time more often active (50 active weeks instead of 12) and more active during these weeks (14 trips per week instead of 2). Note that the decrease of the number of active weeks is not due to a simple seasonal effect: Figure \ref{fig:length_use}) shows that active weeks of regular0 users span a median period of nearly 40 weeks (between the first and the last trip of the year).

\subsection{Comparing the 5-years and 1-year classes}
When comparing the classification obtained here to that found over a single year \cite{vogel}, we note many similarities and a major difference. As in \cite{vogel}, a 'one-off' and a 'week-end' class are found, with similar proportions, as well as six 'regular' classes differing mainly by their intensity of use. The major difference is the 'part-time' class, that represents 3.6\% of users, instead of 29\% for the single year classification (summing their 'intensive and part-time' and 'irregular' classes). This means that those two 1-year classes mostly gathered users that have in fact a \emph{regular} behavior appearing to be 'part-time' because they are observed over a too short period of time. For example, a user starting in September 2011 will appear active only for (at most) 4 months, even if they keep the same activity over the subsequent (unobserved) year. 
Figures \ref{fig:comp_mat_tempa} and \ref{fig:comp_mat_tempb} show that the 'part-time' class gathers users that massively leave the system at the end of the year (nearly 90\% the first year and 70\% the second). Year after year, the 'part-time' class is filled again by users coming from all (previous year) classes, as shown by the numbers in the 'part-time' column in Figures \ref{fig:comp_mat_tempa} and \ref{fig:comp_mat_tempb}. Therefore, this class does not represent a stable behavior of a class of users - people being active every year only 3 months - but the class where users end up for their last year of activity.

\section{Discussion and Conclusion}
As seen in the introduction, there was a lack of in-depth studies on the temporal evolution of long-term bicycle usage, mainly due to the lack of long-term datasets. Thus, we studied the temporal evolutions of year-long V\'elo'v users thanks to a unique dataset spanning over 5 years. After adapting the data, we extended method from \cite{vogel} to characterize temporal patterns and we showed that using a 5-year database corrects the 1-year classification by avoiding the overestimation of part-time users (from 29\% to 3.6\%). This indicates that the seasonal effect in bike usage is much smaller than expected. 

Also, users' yearly activity was organized in three main classes: 6\% of 'one-off' users, with only 3 trips per year; 12\% of 'week-end' users and the rest presenting a regularly distributed activity over the year. We divided the latter class into 6 groups with considerable differences in numbers of trips (from twice a month to twice a day). From these classes, we studied the evolution of activity for longer times and found two main trajectories: About 60\% of the users stay in the system for one year at most and show a low median activity (47 trips); the remaining 40\% of users are more active (median activity of 96 trips in their first year) and remain continuously active for several years (mean time = 2.9 years).

This high proportion of leavers can be explained by many reasons : moving to other towns, buying one's own bike, finding the service unsatisfactory\ldots \ It would be interesting to ascertain the relative proportions of each. Note that considering the number of trips (and not of users) leads to a more stable picture, as the 40\% of stable users perform around 60\% of the annual trips. Their activity is relatively stable, slightly decreasing over the years. We showed that this long-term behavior strongly depends on the user initial class, as fidelity rapidly increases with the number of trips observed the first year. 

On the socio-demographic point of view, stable users are generally older than average users (30 to 40 years old) and live closer to the city center. This result agrees with previous articles socio-demographically characterizing BBS users population (see \cite{beecham,lda_consulting,Ogilvie,shaheen,fuller_et_al,raux_2017}).

The development of temporal study on the evolution of usage and commitment for other BSS in the world would be of high profitability, as it could enable to compare the evolution of trends and socio-demographics in the world.

\section*{Acknowledgement(s)}

JCDecaux and the Grand Lyon Urban Community are gratefully acknowledged for having made V\'elo'v data available to us.

\section*{Disclosure statement}

No potential conflict of interest was reported by the authors.

\section*{Funding}

This research received grants from IXXI and ANR Velinnov ANR-12-SOIN-0001.

\section*{Notes on contributor(s)}

\section*{Nomenclature/Notation}

\bibliographystyle{tfcad}
\bibliography{refs}

\end{document}